\documentclass[epj]{svjour}
\usepackage{graphicx}
\usepackage{amssymb}
\usepackage[latin1]{inputenc} 
\usepackage{url}
\begin{document}

\title{Threshold enhancement in $\eta$ photoproduction from $^2$H and $^4$He}
\author{V.~Hejny\inst{1} \and
J.~Wei{\ss}\inst{2}\and
P.~Achenbach\inst{3} \and
J.~Ahrens\inst{3} \and
J.R.M. Annand\inst{4} \and
R.~Beck\inst{3} \and
M.~Kotulla\inst{2} \and
B.~Krusche\inst{5} \and
V.~Kuhr\inst{6} \and                     
R.~Leukel\inst{3} \and			  
I.J.D. MacGregor\inst{4} \and		 
V.~Metag\inst{2} \and			  
R.~Novotny\inst{2} \and			  
V.~Olmos~de~Le\'on\inst{3} \and		 
F.~Rambo\inst{6} \and			  
A.~Schmidt\inst{3} \and
M.~Schumacher\inst{6} \and
U.~Siodlaczek\inst{7} \and
H.~Str\"oher\inst{1} \and
F.~Wissmann\inst{6} \and
M.~Wolf\inst{2}}

\institute{Institut f\"ur Kernphysik, Forschungszentrum J\"ulich, 
D-52425, J\"ulich, Germany
\and
II. Physikalisches Institut, Universit\"at Gie\ss en, 
D-35392 Gie\ss en, Germany
\and
Institut f\"ur Kernphysik, Universit\"at Mainz, 
D-55099 Mainz, Germany
\and
Department of Physics and Astronomy, University of Glasgow,
Glasgow G128QQ, UK
\and		 
Department of Physics and Astronomy, University of Basel,
Ch-4056 Basel, Switzerland
\and
II. Physikalisches Institut, Universit\"at G\"ottingen, 
D-37073 G\"ottingen, Germany
\and
Physikalisches Institut, Universit\"at T\"ubingen, 
D-72076 T\"ubingen, Germany}

\mail{Bernd.Krusche@unibas.ch}
\date{\today}
\abstract{The photoproduction of $\eta$-mesons from $^2$H and 
$^4\mbox{He}$ has been studied for energies close to the production thresholds.
The experiments were carried out with the tagged photon beam of
the Mainz MAMI accelerator. The $\eta$-mesons were detected via their 
two photon decays with the electromagnetic calorimeter TAPS. Total cross
sections, angular and momentum distributions of the $\eta$-mesons 
have been determined for both reactions. The total cross
sections in the threshold region show a large enhancement over the predictions
of a participant - spectator model, indicating significant final state
interaction effects. The results are compared to recent model calculations
taking into account nucleon-nucleon and nucleon-$\eta$ final state interaction
effects on different levels of sophistication.}

\maketitle
\section{Introduction}
\label{sec:intro}
The interaction of mesons with nucleons and nuclei is one of the central issues
in strong interaction physics.
Detailed experimental and theoretical studies of the pion - nucleon and 
pion - nucleus interaction have largely contributed to this field. However,
much less is known in case of the heavier pseudoscalar mesons, the $\eta$- and
$\eta'$-mesons, which due to their short lifetimes are not available as
beams. In particular the $\eta$-meson has recently attracted much interest. 
At small meson - nucleon relative momenta the situation for $\eta$-mesons and 
pions is very different. This is due to the fact that an s-wave nucleon
resonance, the S$_{11}$(1535), which is located very closely to the 
production threshold for $\eta$-mesons, couples strongly to the $N\eta$-system.
As a 
consequence reactions like photoproduction of $\eta$-mesons are completely 
dominated by the excitation of this resonance \cite{Krusche95a}. 
This feature of the $N\eta$-system has consequences for the
interaction of $\eta$-mesons with nuclei. It has even been argued \cite{Ueda} 
that due to the $N\pi$-$N\eta$ interaction in the $S_{11}$-state the 
$NN\eta$-system might be bound.

Most information of the $N\eta$-interaction, in particular the $\eta$-nucleon
scattering length, has been extracted from coupled channel analyses of pion and
photon induced meson production reactions. Already in 1985 Bhalerao and Liu 
\cite{Bhalerao} found an attractive s-wave $N\eta$ interaction with a 
scattering length of $a_{\eta}=(0.27 + 0.22i)$ fm from such an analysis. 
Shortly afterwards, Liu and Haider \cite{Liu} pointed out that for nuclei 
with $A>10$ this interaction might lead to the  formation of bound 
$\eta$-nucleus states which they termed $\eta -mesic$ nuclei. Experimental 
evidence for such 'heavy' $\eta$-mesic nuclei was sought in different 
reactions \cite{Chrien,Johnson,Sokol}, but so far with non-conclusive results.
   
Substantial progress has been achieved during the last few years in the
experimental study of $\eta$-production reactions. Precise experiments using
hadron induced reactions (e.g. at CELSIUS, COSY, SATURNE)
\cite{Mayer96,Plouin90,Calen96,Calen98,Willis,Hibou,Beligeri,Smyrski} or photon
induced reactions (e.g. at ELSA, GRAAL, JLAB, KEK, MAMI) 
\cite{Krusche95a,Krusche95b,Roebig,Hoffmann97,Ajaka98,Yorita,Thompson,Weiss01,Renard}
on the free nucleon and on nuclei have vastly improved the available data 
base. 
The new results for the elementary reactions on the nucleon, 
in particular for the photon induced reaction \cite{Krusche95a}, have prompted
several groups to re-analyse the $\eta N$ interaction 
\cite{Green99,Wilkin93,Sauer95,Arima92,Batinic96,Abaev96}.
They found $N\eta$-scattering lengths with real parts in the range 0.5 -1. fm,
considerably larger than the original result from Bhalerao and Liu 
\cite{Bhalerao}. Furthermore, strong threshold enhancements were observed
in hadron induced $\eta$-production reactions.
In the following the possible existence of $\eta$-nucleus bound states for the
light nuclei $^2\mbox{H}$, $^3\mbox{H}$, $^3\mbox{He}$ and $^4\mbox{He}$ was
intensely discussed in the literature
\cite{Willis,Wilkin93,Rakityanski,Green96,Scoccola,Shevchenko,Grishina,Garcilazo},
but there is still some controversy regarding the necessary strength
of the $\eta N$-interaction.

The investigation of $\eta$-production reactions from light nuclei in the
threshold region is a very useful tool for the study of eta-nucleus
interactions. The idea is, that final state interaction (FSI) of the
$\eta$-meson with the nucleus, in particular the formation of quasibound 
states in the vicinity of the production threshold, will give rise to an
enhancement of the cross section relative to the expectation for phase space
behavior. Such deviations can also arise from FSI effects between
the nucleons, which are not related to the $\eta N$-interaction, so that
the interpretation of the results needs at least reliable model calculations
which account for these trivial effects. 
During the past five years $\eta$-production near threshold was 
intensively investigated in the reactions:
$pp\rightarrow pp\eta$ \cite{Calen96}, $np\rightarrow d\eta$ 
\cite{Plouin90,Calen98}, $pd\rightarrow\eta  ^3\mbox{He}$
\cite{Mayer96}, $dd\rightarrow \eta ^4\mbox{He}$ \cite{Willis}, 
and $pd\rightarrow pd\eta$ \cite{Hibou}. All reactions show more or less 
pronounced threshold enhancements. 

Possible $\eta$-nucleus FSI effects should show up independently of the 
initial state of the reaction. Photoproduction of
$\eta$-mesons from light nuclei is a clean tool for the preparation of 
the $\eta$ - nucleus final state with small relative momenta, but due to the 
small electromagnetic cross sections no sensitive threshold 
measurements have been reported until now. Previously, $\eta$-photoproduction
from the deuteron was studied in two experiments carried out at the MAMI and
ELSA accelerators \cite{Krusche95b,Hoffmann97}, which aimed at the extraction 
of the isospin composition of the electromagnetic excitation of the 
S$_{11}$(1535) resonance. Both experiments did not reach the statistical 
accuracy necessary for a sensitive search for threshold effects. In the 
present paper we report the results from a new measurement of the 
$d(\gamma ,\eta)X$ reaction with an improvement in counting statistics by 
one order of magnitude compared to ref. \cite{Krusche95b} and the first 
measurement of the threshold behavior of $\eta$-photoproduction from $^4$He. 
The data have been measured as part of a systematic investigation of 
$\eta$-photoproduction from light nuclei. The results from the 
$^4$He measurement concerning the isospin structure of the
S$_{11}$(1535) excitation have been published in ref. \cite{Hejny99}. 

The excitation of the S$_{11}(1535)$ resonance, which dominates $\eta$-threshold
production, involves an electromagnetic spin-flip transition which is
predominantly of isovector nature. The isoscalar admixture in the amplitude
contributes only at the 10\% level 
\cite{Krusche95b,Hoffmann97,Weiss01,Hejny99}. 
Consequently, coherent $\eta$-photo\-pro\-duction, where the nucleus remains 
in the ground state, is suppressed for the J=1, I=0 deuteron
\cite{Hoffmann97,Weiss01}
and almost entirely forbidden for the I=J=0 nucleus $^4\mbox{He}$
\cite{Hejny99}. As a result
inclusive reactions are dominated by incoherent breakup processes.
Nevertheless, close to threshold, energy conservation requires that the
kinematical conditions must be similar to the coherent process. The
relative momentum of the $np$-pair in the $\gamma d\rightarrow\eta np$
reaction must be small so that nucleon - nucleon FSI will certainly be
important.  

\section{Experiment and analysis}

Eta photoproduction from the deuteron and $^4\mbox{He}$ was measured close
to the production thresholds at E$_\gamma$=627 MeV and 587 MeV, respectively.
The experiments were carried out at the Glasgow tagged
photon facility \cite{tagger} installed at the Mainz Microtron (MAMI) 
\cite{mami} with the electromagnetic calorimeter TAPS \cite{taps,Gabler}.  
A detailed description of the setup and the analysis is given in \cite{Hejny99}.
Here we summarize only the most relevant aspects of the data analysis. The
identification of photons was done with a time-of-flight measurement with
typically 500 ps resolution (FWHM), the pulse-shape discrimination capability of
the BaF$_2$-scintillators, and the plastic scintillators mounted as charged
particle veto detectors in front of the BaF$_2$-crystals. The combination of 
these
methods assures that contaminations from particles are below the 1 \% level.
The $\eta$-mesons were then identified with a standard invariant mass analysis
of coincident photon pairs using:
\begin{equation}
  \label{eq:etot}
  m_{\mathrm{inv}}^2  = ( E_1 + E_2 )^2 - (\vec{p}_1 + \vec{p}_2)^2 
          =  2 E_1 E_2 (1-\cos \Phi_{12}) 
\end{equation}
where $E_i$ denotes the energies, $\vec{p}_i$ the momenta of the two photons
and $\Phi_{12}$ the relative angle between them. Background in the invariant
mass spectra originated only from events of multiple $\pi^o$-photoproduction
\cite{Hejny99} when two photons from different pions were observed and the other
photons escaped detection due to the limited solid angle coverage of the
detector. However, as shown with Monte Carlo simulations \cite{Hejny99}, in the
invariant mass spectra this background levels off under the low energy tail of
the $\eta$-peak and can thus be removed with a cut on the invariant mass. The
fact that the excitation functions for $\eta$-photoproduction (see fig.
\ref{fig1}) are consistent with zero below the respective $\eta$-production
thresholds is further evidence that this background was completely removed by
the invariant mass cut.

\begin{figure}[t]
\resizebox{0.5\textwidth}{!}{
\includegraphics{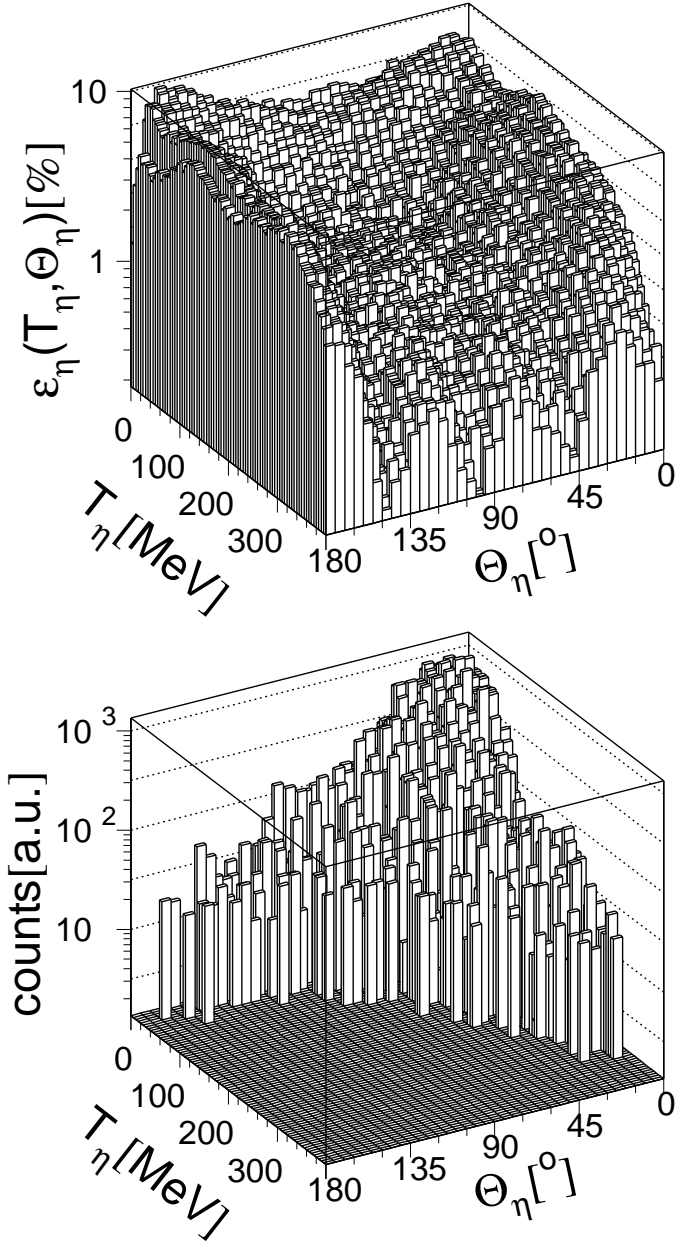}
}
\caption{Detector acceptance for $\eta$-mesons as function of the laboratory 
polar angle and the laboratory kinetic energy of the mesons. Upper 
part: efficiency calculated with Monte Carlo simulation. Lower part: measured 
distribution of $\eta$-mesons for incident photon energies 660 - 680 MeV.}
\label{fig0}
\end{figure}

The detection efficiency of the TAPS-detector for 2$\gamma$-decays of
$\eta$-mesons was determined with Monte Carlo simulations based on the
GEANT-code \cite{GEANT}. It was previously demonstrated \cite{Gabler} that
this simulations reproduce the photon response of the detector very precisely.
The detection efficiency $\epsilon_{\eta} (\Theta_{\eta} ,T_{\eta})$ depends
only on the laboratory polar angle $\Theta_{\eta}$ and the laboratory kinetic 
energy $T_{\eta}$ of the $\eta$-mesons, which were both measured in the 
experiment.
The efficiency was not large due to the limited solid angle coverage. However, 
in the investigated range of incident photon energies it was not zero for any
kinematically possible combination of $\Theta_{\eta}$ and $T_{\eta}$, because
always some fraction of the decay photon pairs was in the acceptance of the
calorimeter. This is demonstrated in fig. \ref{fig0} where for one range of 
incident photon energies the detector efficiency and the measured distribution
of $\eta$-mesons are compared. Therefore no model assumptions were needed for 
the determination of
the efficiency correction. It was obtained from the Monte Carlo simulations
as function of $\Theta_{\eta}$ and $T_{\eta}$ and corrected event-by-event.
Finite resolution effects were accounted for in an iterative procedure: in a
first step $\epsilon_{\eta}$ was simulated with an isotropic angular and energy
distribution of the $\eta$-mesons. This efficiency was then used to correct the
data and in the following the distributions extracted from the data were used as
input for the simulations. 
This procedure converged after 1 - 2 iterations. The
efficiency correction for the total cross section determined this way varies
smoothly between 6.5 - 5 \% for the first 50 MeV of incident photon energies
above the production thresholds. The systematic uncertainty of the efficiency is
below 5 \%. This uncertainty has been investigated in detail for the $\gamma
p\rightarrow p\eta$ reaction \cite{Krusche95c}. In this case the 3$\pi^o$-decay
of the $\eta$-meson was analysed in addition to the 2$\gamma$-decay.
The two decay channels have very different detection efficiencies both on an 
absolute scale as well as for the energy dependence. 
The energy dependence of the excitation
functions extracted from the two decay channels agreed within the statistical
uncertainties of a few per cent. The absolute value of the cross section ratio
was consistent with previous measurements of the ratio of the decay widths
$\Gamma_{\eta\rightarrow 2\gamma}/\Gamma_{\eta\rightarrow 3\pi^o}$ but had even
a smaller uncertainty.

The absolute normalization of the cross sections was obtained from the target
thicknesses and the intensity of the photon beam. The latter was determined by
counting the deflected electrons in the tagging spectrometer and measuring the
tagging efficiency (i.e. the fraction of the correlated photons which pass
through the photon collimator) by moving a BGO-detector into the photon beam at
reduced intensity. The uncertainty of this normalization is estimated as 3 \%.

The threshold behavior of $\eta$-photoproduction was studied in inclusive
experiments where only the $\eta$-mesons were detected. Exclusive measurements
of quasifree $\eta$-pho\-to\-pro\-duction from the neutron via coincident detection 
of recoil nucleons at higher incident photon energies \cite{Hejny99} and an 
exclusive measurement of coherent $\eta$-photoproduction from the deuteron 
via coincident detection of the recoil deuterons \cite{Weiss01} are reported 
elsewhere.
    
\section{Results and discussion}
\begin{figure}[t]
\resizebox{0.483\textwidth}{!}{
\includegraphics{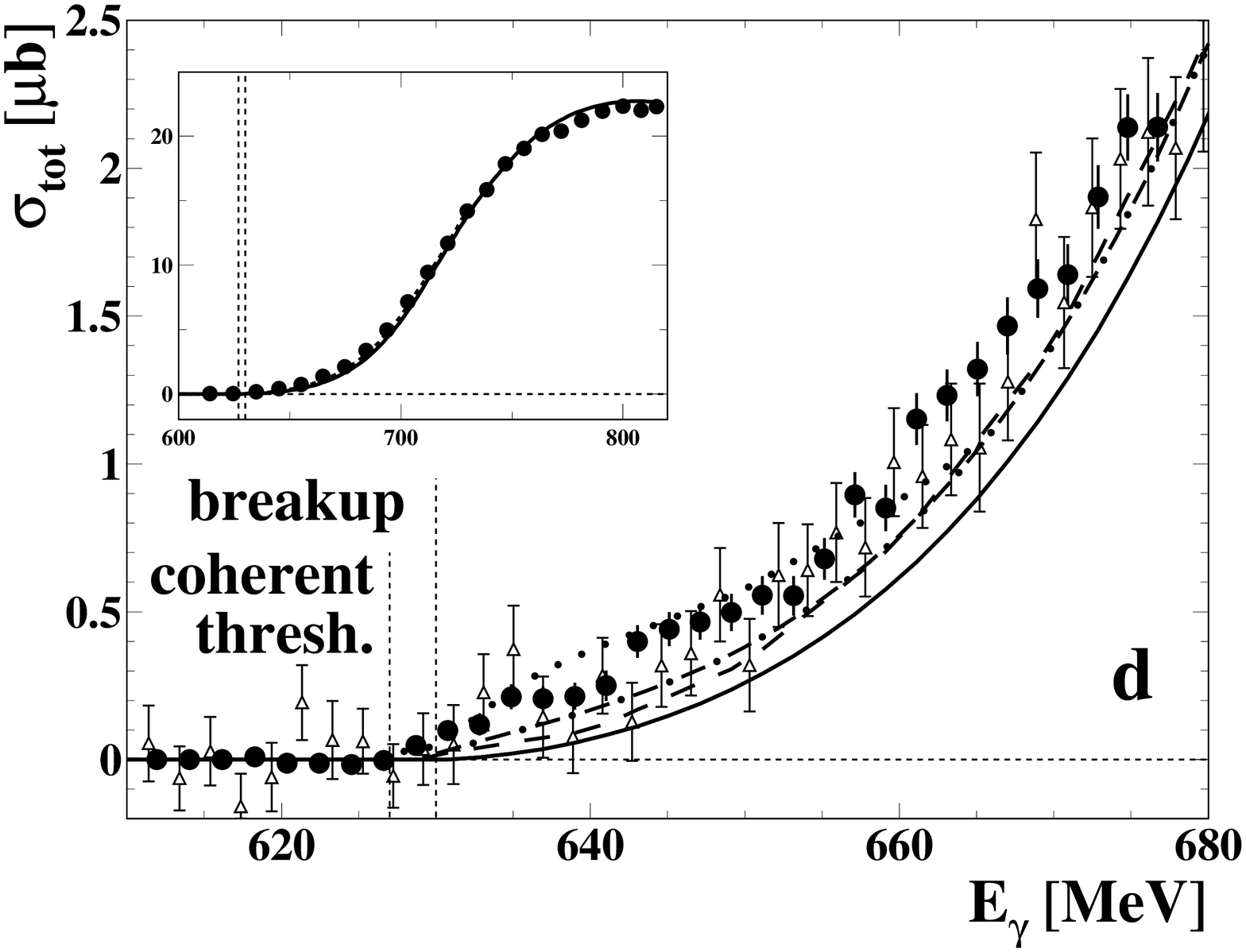}
}
\resizebox{0.483\textwidth}{!}{
\includegraphics{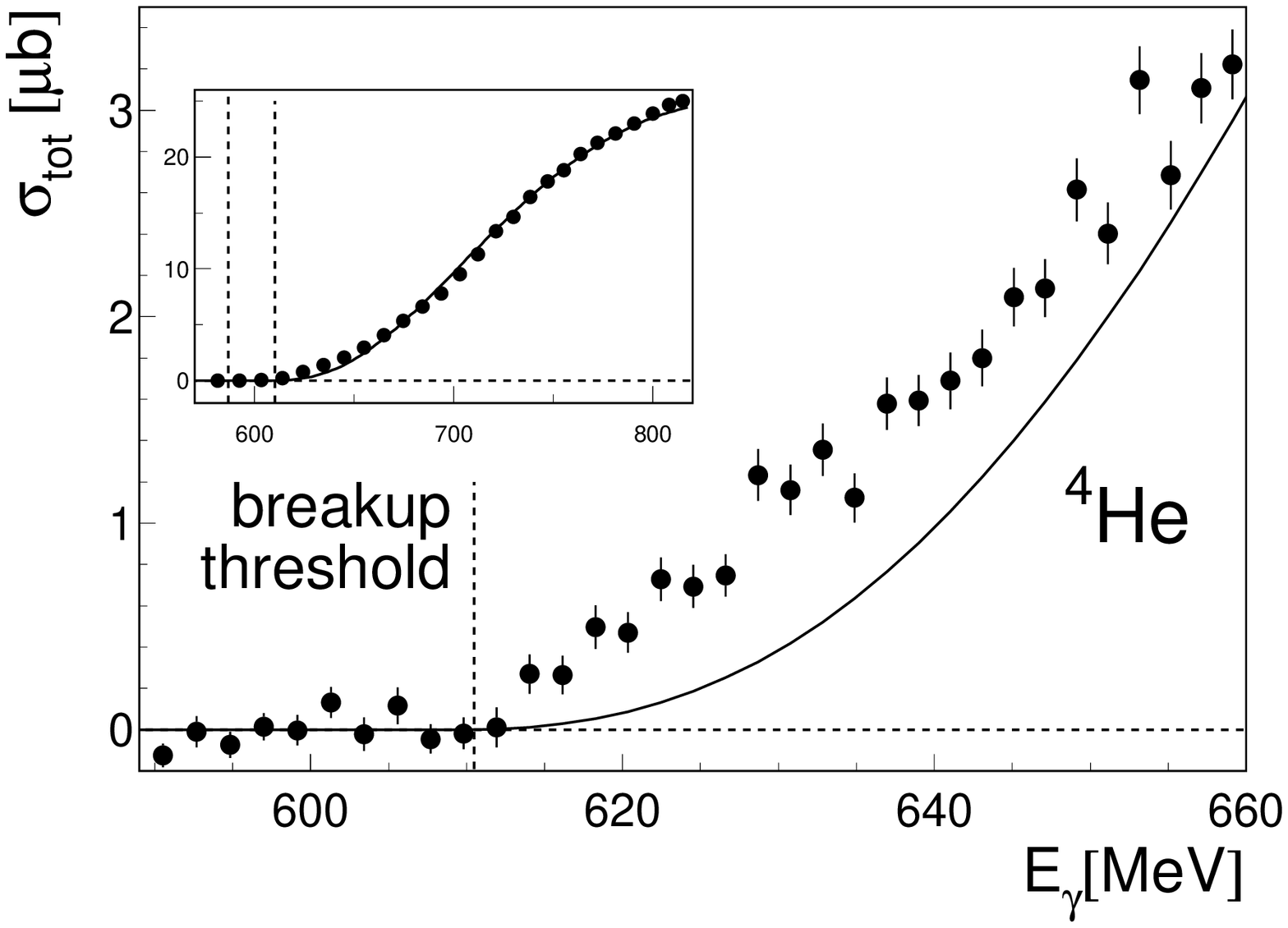}
}
\caption{Total cross sections for $\eta$ photoproduction from 
the deuteron (upper part) and $^4\mbox{He}$ (bottom part). 
Main plots: threshold region; inserts: full measured energy range.
The dashed vertical lines indicate the coherent and quasifree thresholds. 
Solid curves: impulse approximation. 
For deuterium:
open triangles: data from ref. \protect\cite{Krusche95b} 
dash-dotted curve: model with $NN$ FSI 
(Sibirtsev et al. \protect\cite{Sibirtsev}),
dashed curve: model with first order $NN$ and $\eta N$ FSI 
(Fix et al. \protect\cite{Fix97}),
dotted curve: three body model 
(Fix et al. \protect\cite{Fix01}) 
}
\label{fig1}
\end{figure}

The measured total cross sections are shown in fig. \ref{fig1}.
In the helium case, the data are consistent with zero at incident photon 
energies below the breakup threshold. This agrees with the expectation that 
the cross section of the coherent reaction is negligible. A small coherent 
contribution might be visible for the deuteron between the coherent and 
breakup thresholds. For deuterium also the older data \cite{Krusche95b} with 
larger statistical
errors is shown. The two data sets are in good agreement within their
statistical uncertainties. Very close to threshold both data sets might exhibit
a structure, which however is not statistically significant.

\begin{figure}[t]
\resizebox{0.483\textwidth}{!}{
\includegraphics{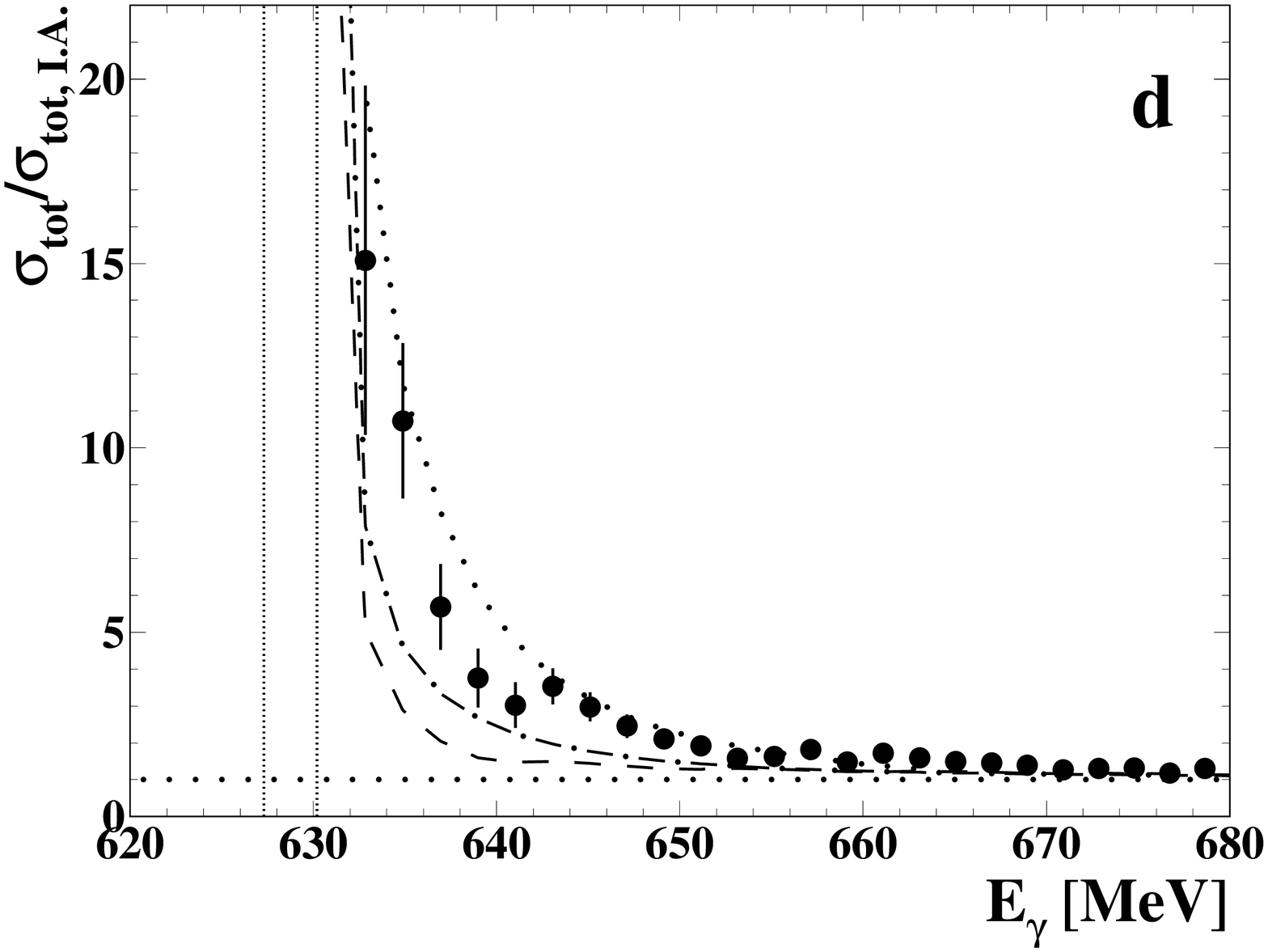}}
\resizebox{0.481\textwidth}{!}{
\includegraphics{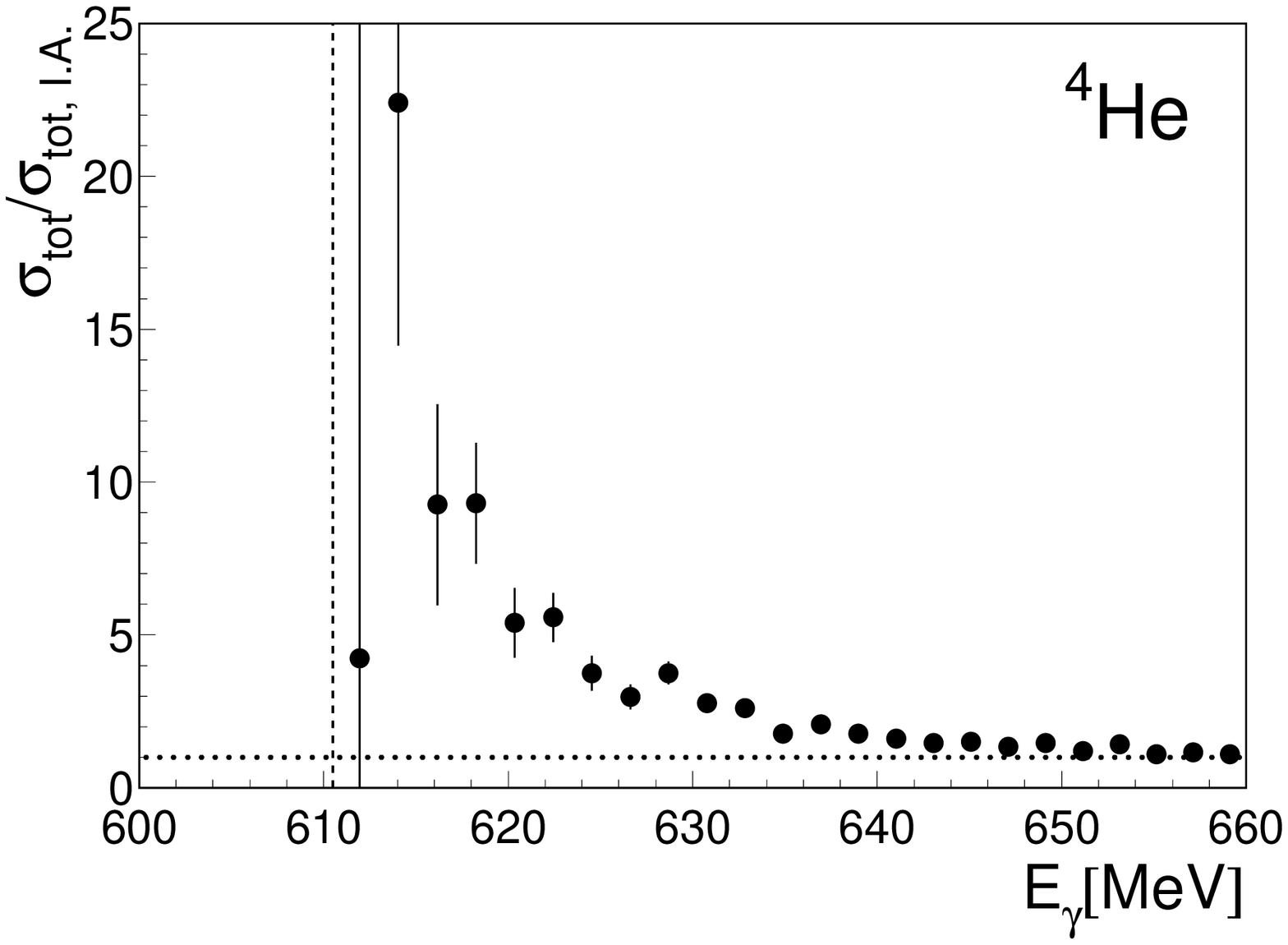}
}
\caption{Experimental data and model calculations normalized to the impulse 
approximation. The curves correspond to the same models as in fig. 
\protect\ref{fig1} and are also normalized to the impulse approximation.}
\label{fig2}
\end{figure}

In a first step the data are compared with a simple participant - spectator 
model in the impulse approximation. This model assumes that the ratio 
$\sigma_n /\sigma_p$ of the 
cross sections for $\eta$-photo\-pro\-duction from the proton and the neutron 
is constant over the investigated energy range. A Breit-Wigner fit to the 
proton data \cite{Krusche95a,Krusche95b} was folded with the momentum 
distribution of the bound nucleons and normalized to the nuclear data
by a constant factor  representing the cross section ratio $\sigma_n /\sigma_p$.
The momentum distributions of the nucleons were generated from the 
deuteron \cite{dwave} and helium wave functions. The latter was calculated 
from the measured $^4$He charge form factor \cite{hewave}. 
The proton form factor was unfolded and the proton spatial 
density was generated via a Fourier transformation. The spatial wave function
was taken as square root of the density distribution and the wave function in
momentum space was calculated from a second Fourier transformation. This is
based on the assumptions that $^4$He is purely s-wave and protons and neutrons
have the same distributions. As already reported in \cite{Krusche95b,Hejny99} 
good agreement between model and data over a large energy range is 
obtained in both cases under the assumption that 
$\sigma_n /\sigma_p\approx 0.66$. This result also agrees with exclusive
experiments where the cross section ratio is determined via coincident detection
of the $\eta$-meson and the recoil nucleon \cite{Hoffmann97,Hejny99}.
However, in the threshold region the data are significantly underestimated by 
this model. For both nuclei the measured cross sections rise much faster
than the predictions at energies above the breakup threshold. The size of the 
effects is more visible in fig. \ref{fig2} where the data are divided by the
impulse approximation calculation.


\begin{figure*}[t]
\begin{center}
\resizebox{0.75\textwidth}{!}{
\includegraphics{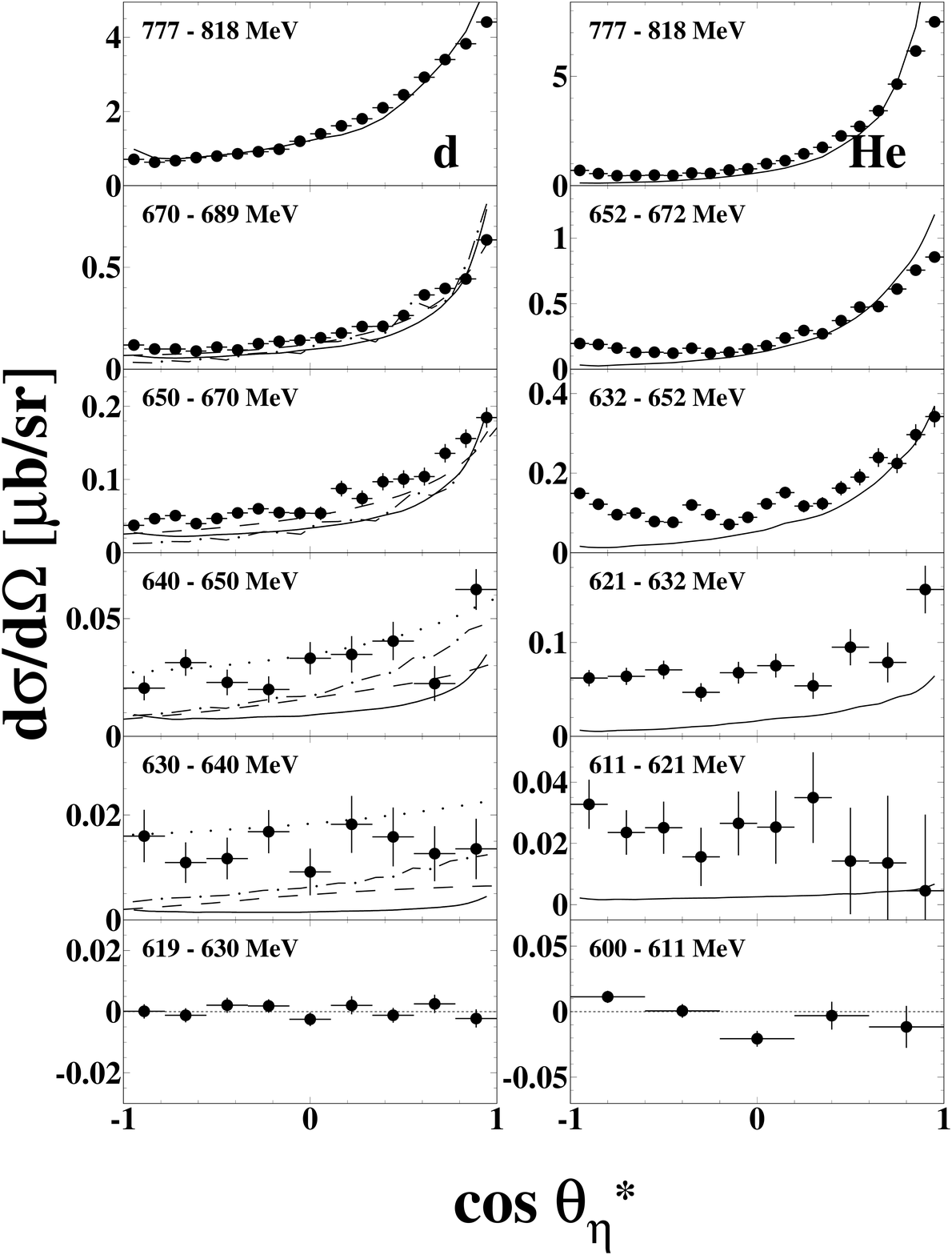}
}
\caption{Angular distributions of the $\eta$ mesons in the photon-nucleus 
cm system. The data are divided in 5 energy bins above threshold and one bin 
below threshold which shows that background is negligible. 
Left hand side: deuterium, right hand side: helium.
Solid curves: impulse approximation.
The dashed, dash-dotted and dotted curves for deuterium
are the model predictions from \protect\cite{Fix97,Sibirtsev,Fix01} as in 
figs. \ref{fig1}, \ref{fig2}.}
\label{fig3}
\end{center}
\end{figure*}
\begin{figure*}[t]
\begin{center}
\resizebox{0.75\textwidth}{!}{
\includegraphics{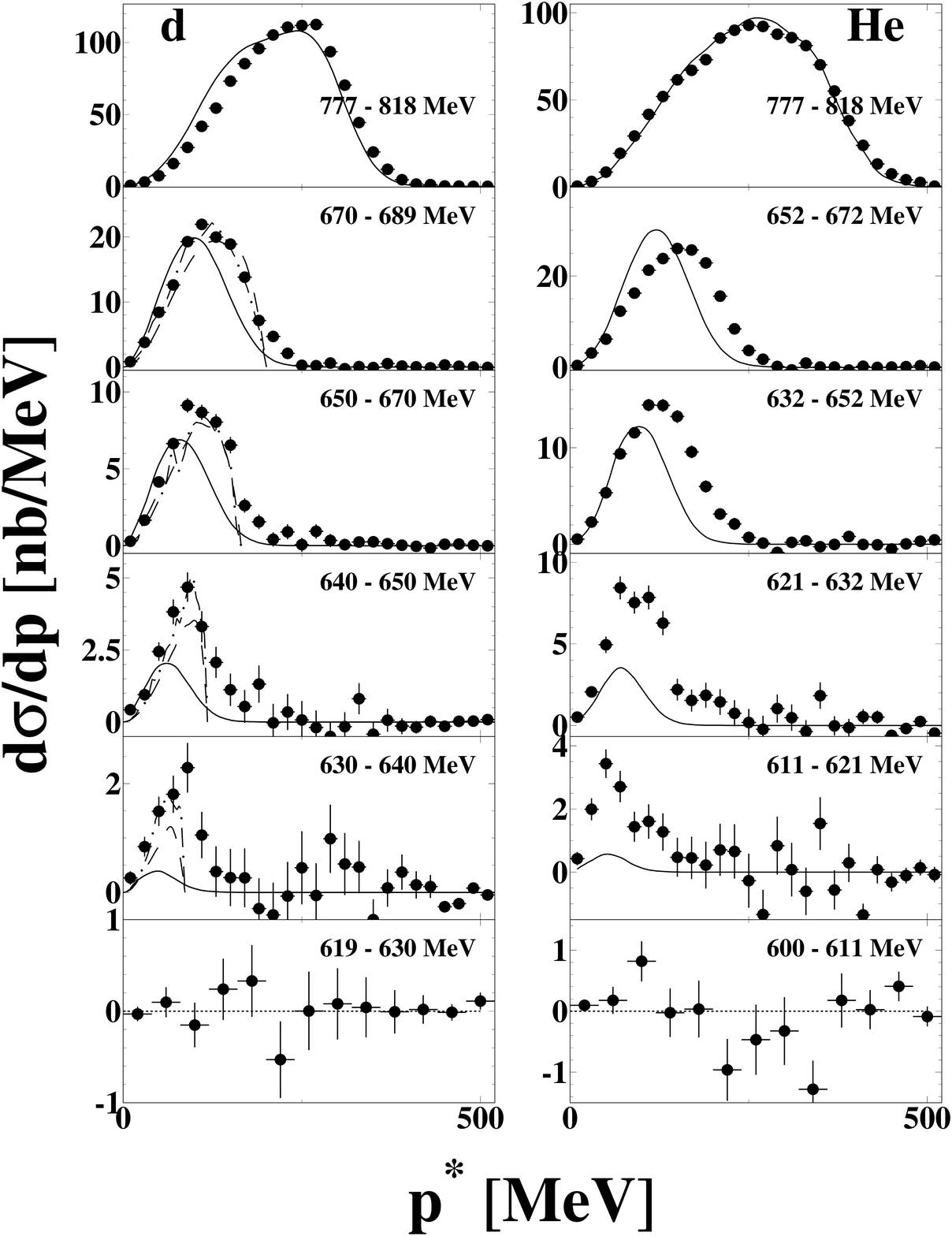}
}
\caption{Momentum distributions of the $\eta$-meson in the photon-nucleus
cm frame. Solid curves: participant - spectator model folded with the 
experimental resolution. The dashed and dash-dotted curves for deuterium
are the model predictions from \protect\cite{Fix97,Sibirtsev} as in figs.
\ref{fig1}, \ref{fig2}.}
\label{fig4}
\end{center}
\end{figure*}

The large enhancement over the impulse approximation at threshold is 
evidence for substantial FSI effects, which are not included in the model.
In the deuteron case model calculations from different groups,
which include nucleon - nucleon and nucleon - $\eta$ FSI effects, have recently
become available and are compared to the data in figs. \ref{fig1}, \ref{fig2}. 
Sibirtsev et al. \cite{Sibirtsev} have studied in detail the influence of $NN$ 
FSI on the cross section. In agreement with our results they reproduce the
cross section in impulse approximation at higher incident photon energies but
find a significant contribution of $NN$ FSI in the threshold region.
Interestingly, in contrast to hadron induced $\eta$-production, the $NN$ FSI 
effects turn out to be rather insensitive to details of the $NN$ potential. 
However, already a comparison \cite{Sibirtsev} to our older low 
statistics data \cite{Krusche95b} suggests that
the effects of $NN$ FSI are not large enough to explain the threshold
enhancement. This result is corroborated by 
the new data (see figs. \ref{fig1}, \ref{fig2}, dash-dotted
curves). This is an indication that $N\eta$ FSI must be important.
Fix and Arenh\"ovel have calculated FSI effects from $NN$ and $N\eta$
rescattering \cite{Fix97}. They also find a significant contribution from
$NN$ rescattering but less important effects from $N\eta$-rescattering. However,
their final result (see figs. \ref{fig1}, \ref{fig2}, dashed curves) still 
underestimates the threshold data. 
Very recently the same authors have presented a Faddeev type three-body
calculation of the $NN\eta$-system with separable two-body interactions
\cite{Fix01}. In this calculation they find that the $\eta NN$ three-body 
aspects are important in the threshold region. The result of this model
(see figs. \ref{fig1}, \ref{fig2}, dotted curves) is close to the data
in the threshold region but seems to underestimate it at higher incident
photon energies. However, the authors mention that in order to treat explicitly
the three-body aspects they have simplified some other model ingredients.
In particular, they have neglected the d-wave component of the deuteron wave
function and the short range $NN$-repulsion. Both effects are expected to reduce
the pure s-wave cross section in the threshold region. The agreement with the
data therefore must be taken with some care.  

Model predictions including FSI effects are not yet available for the
$^4\mbox{He}$ case. Here, the situation is much more complicated since 
calculations must account for the various exit channels $pt$,
$n^{3}\mbox{He}$, $dd$, $pnd$, $2p2n$ with the corresponding kinematical 
thresholds. The successive opening of the thresholds could also contribute to 
the observed threshold enhancement, although the shape of the relative 
enhancement looks quite similar for the deuteron and helium case.

Further clues to the nature of the threshold enhancement can be expected from
the angular and momentum distributions of the $\eta$-mesons. In figs.
\ref{fig3},\ref{fig4} they are summarized and compared to the results of the
models. The angular
distributions (see fig. \ref{fig3}) in the photon - nucleus cm system
agree quite well with the impulse approximation at the highest incident 
photon energies close to the position of the S$_{11}(1535)$ resonance.
In the case of deuterium the agreement is reasonable down to photon energies
around 650 MeV, but for helium an excess at backward angles develops already at
higher energies. As expected, effects beyond the plane wave approximation
are larger for the strongly bound helium nucleus. Very close to threshold the 
measured angular distributions are almost isotropic.
The shape change of the angular distributions as a function of the incident 
photon energy is partly due to the reaction kinematics.
As already mentioned, the dominant production processes of the $\eta$-mesons
are breakup reactions where the meson is produced on an individual nucleon that
is subsequently knocked out of the nucleus. The elementary reaction on a nucleon
through the S$_{11}$ excitation is almost pure s-wave. Consequently, as long as
Fermi motion is neglected the angular distributions are expected to be isotropic
in the cm frame of the photon and a {\it nucleon} at rest. This behavior has 
been demonstrated for the deuteron \cite{Krusche95b} and
is also largely true for helium, although in this case some enhancement at
backward angles is visible at all incident energies \cite{Hejny99}. The angular
distribution shown here refer to the slower photon - {\it nucleus} rest frame
so that they appear forward boosted. Since at low incident photon energies,
energy conservation enforces an asymmetric selection of Fermi momenta
antiparallel to the incident photon momentum, the 'true' average cm
frame has a velocity somewhere in between the two extremes and moves closer to
the photon - nucleus frame. But this kinematical effect, which is included
in the impulse approximation, does not explain the observed change in shape
close to threshold. This is again an indication for significant FSI effects. 
However, for the photon energy range 630 - 640 MeV also the model results 
for deuterium with $NN$ \cite{Sibirtsev} or $NN$ and 
$\eta N$ \cite{Fix97} FSI effects predict a rise of the cross section from 
backward to forward angles by a factor of 3 - 4, which clearly does not agree
with the observed isotropy. Only the three-body calculation \cite{Fix01}
shows a flatter behavior.
 
\begin{figure}[h]
\resizebox{0.5\textwidth}{!}{
\includegraphics{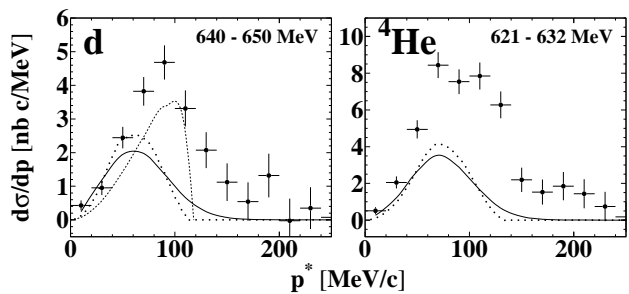}
}
\caption{Momentum distributions of the $\eta$-meson in the photon-nucleus
cm frame. Solid curves: participant - spectator model folded with the 
experimental resolution. Dotted curves: participant - spectator approximation
without finite resolution effects. Dashed curve: calculations from ref. 
\protect\cite{Fix97}.}
\label{fig5}
\end{figure}

The momentum distributions in the photon - nucleus cm frame are compared to
model predictions in fig. \ref{fig4}. Again the shape of the 
distributions is nicely reproduced by the participant - spectator approximation
at high incident photon energies but not in the threshold region. 
The models with final state interaction from \cite{Fix97,Sibirtsev} 
(predictions from ref. \cite{Fix01} are not available) reproduce the deuterium
data much better at lower energies. Here one should note, that Monte Carlo 
simulations of the detector response indicate that in contrast to the angular 
distributions the momentum distributions are significantly affected by the 
finite detector resolution. The plane wave approximation was therefore folded 
with the detector response. This could not be done for the model predictions so
that most of the disagreement between models and data at high momenta
is artificial.  
The size of the effect is shown in fig. \ref{fig5} where for one energy range
folded and unfolded participant - spectator model predictions are compared to 
the data. The resolution effects somewhat broaden the momentum distributions 
but do not shift them. The model predictions from \cite{Fix97} are not folded 
with the experimental resolution which explains the sharp cutoff at large 
momenta.

\section{Conclusions}

It has been shown, that in $\eta$ photoproduction from $^2$H and $^4$He
the total cross section at threshold is strongly enhanced with respect
to an impulse approximation calculation. In the case of deuterium, where
calculations are available, models which include $NN$ and $\eta N$ FSI effects
account for only part of this enhancement. Only a recent three-body
calculation of the $NN\eta$-system is in better agreement with the threshold
data but fails at higher incident photon energies.
Both the angular and the momentum distributions of the mesons 
show evidence for significant final state interaction.
The enhancement bears some similarities to the threshold behavior of
$\eta$-production found in hadron induced reactions which has been taken as
tentative evidence for the formation of $\eta$-nucleus quasibound states.
However, no firm conclusion as to FSI being strong enough to give rise to
quasi-bound states can be drawn at this stage.

~\\
{\bf Acknowledgments}

We wish to acknowledge the outstanding support of the accelerator group of
MAMI, as well as many other scientists and technicians of the Institut f\"ur
Kernphysik at the University of Mainz. Illuminating discussions on the
theoretical interpretation of the experimental results with H. Arenh\"ovel, A.
Fix, A. Sibirtsev and C. Wilkin are gratefully acknowledged.  
This work was supported by Deutsche
Forschungsgemeinschaft (SFB 201), Bundesministerium f\"ur Bildung und
Forschung (BMBF, Contract No. 06 GI 475(3) I) the UK Engineering and
Physical Science Research Council and the Swiss National Science Foundation.

\end{document}